\documentclass[preprint]{aastex}

\shorttitle{NGC 5128 GCS}
\shortauthors{Woodley et al.}

\begin{document}

\title{Radial Velocities of Newly Discovered Globular Clusters in NGC 5128}

\author{Kristin A.~Woodley\altaffilmark{1}}
\affil{Department of Physics \& Astronomy, McMaster University,
  Hamilton ON L8S 4M1}
\email{woodleka@physics.mcmaster.ca}

\author{William E.~Harris\altaffilmark{1}}
\affil{Department of Physics \& Astronomy, McMaster University,
    Hamilton ON L8S 4M1 }
\email{harris@physics.mcmaster.ca}

\author{Gretchen L.~H.~Harris}
\affil{Department of Physics, University of Waterloo, Waterloo ON N2L 3G1}
\email{glharris@astro.uwaterloo.ca}

\altaffiltext{1}{Visiting Observer,
Cerro Tololo Interamerican Observatory, Operated by the Association 
of Universities for Research in Astronomy, Inc. (AURA) under
cooperative agreement with the National Science Foundation.}

\begin{abstract}
We present radial velocity measurements for 74 globular clusters
(GCs) in the nearby giant elliptical NGC 5128, of which 31 are newly
discovered clusters.  All the GC candidates were taken from the list of
possible new clusters given in the \citet{hhgII} photometric survey.
In addition to the newly confirmed clusters, we identified
24 definite foreground stars and 31 probable
background galaxies.  From a combined list of 299 known GCs in
NGC 5128 with measured radial velocities and metallicity-sensitive
$(C-T_1)$ photometric indices, we construct a new
metallicity distribution function (MDF) for the cluster system.
The MDF shows an approximately bimodal form, with centroids at
$\rm{[Fe/H]} = -1.46$ and $-0.53$, and with nearly
equal numbers of metal-poor and metal-rich clusters in the two modes.
However, there are many intermediate-color clusters in the distribution,
and the fainter clusters tend to have a higher proportion of red clusters.  These features
of the MDF may indicate a widespread age range within the cluster
system as well as an intrinsically broad metallicity spread.  
\end{abstract}


\section{Introduction}

Globular clusters (GCs) are simple stellar populations (single
age, single metallicity) that help to trace
the formation history of their host galaxy.  GCs are also
powerful tools in revealing the halo kinematics and dynamics, 
if large samples of globular cluster radial velocities can
be obtained.

NGC 5128, the central galaxy in the Centaurus cluster 3.9 Mpc
away, and the nearest readily accessible giant E galaxy, 
provides a unique and exciting possibility to study the
dynamics and formation history through its globular
cluster system (GCS).  However, finding its clusters individually
is a difficult job:  at its moderately low Galactic latitude
of $19^o$, field contamination by both foreground stars and background
galaxies is significant, and the GCs cannot be
distinguished by only image morphology, object size, or color.  Thus except for
the very expensive and time-consuming approach of resolving the GCs 
into individual stars by (say) HST imaging, radial
velocity determinations provide the only definitive method for classifying
a candidate object in the NGC 5128 field as a genuine GC.  

NGC 5128 has a statistically determined total population of
$980 \pm 120$ globular clusters, giving it a low specific frequency 
$S_N = 1.4 \pm 0.2$ (Harris, Harris, \& Geisler
2004, hereafter HHG04).  An $S_N$ value this low (in the range
occupied by large spiral galaxies) is suggestive of a merger
origin \citep{harris}; other indicators of mergers
are the faint isophotal shells in the halo of the galaxy \citep[e.g.][]{peng02}, and the 
activity in the inner 5 kpc region indicated by the
visible dust lane, the recent star formation, and the presence of gas.
The GCS, with its wide mixture of cluster metallicities and perhaps ages,
should provide many additional clues to the formation history surrounding this
giant elliptical.  An excellent starting point for any such discussion is to have as
large as possible a list of individual, certainly identified GCs through
radial velocity measurements.

Previous work has led to a list of 215 confirmed GCs in NGC 5128 with radial 
velocity measurements (Peng, Ford, \& Freeman
2004, hereafter PFF04).  Many hundreds more clusters remain to be found.
In this paper, the first of a new series of studies of the GCS in this
unique galaxy, we present new velocity measurements and the
discovery of 31 new GCs.

\section{Data Reduction and Velocity Measurements}

The wide-field photometry study of the NGC 5128 field in
the Washington $CMT_1$ system by \citet{hI} provides a list of
327 GC candidates having globular-cluster-like 
colors and slightly nonstellar morphology.  This list provides the basis for our
new radial velocity survey.

\subsection{Observations and Data Reduction}

Observations were taken at the Cerro Tololo Inter-American Observatory
(CTIO) 4m telescope in April 2004.  The Hydra multi-object fibre-fed
spectrograph, with $40'$ diameter field of view, was used with the KPGL1 
grating and $2 \times$ binning in dispersion, giving a spectral resolution element
of approximately 3.5 binned pixels (FWHM).  The spectral coverage was 
$\lambda3900$~$\AA - \lambda5300$~$\AA$, extending from Ca II H+K at the blue
end to the Mgb features at the red end at 1.18 \AA/px.  Of our three assigned nights,
two were lost to clouds, and the third had
poor seeing throughout ($2'' - 3''$). Nevertheless, we successfully
observed two Hydra field configurations, both centered on the galaxy.
Total exposure times were 
$6 \times 1800$ sec for the first field and $7 \times 1800$ sec for the second.

The data reduction was performed through the {\it dohydra} package
within IRAF\footnote{IRAF is distributed by the National Optical Astronomy Observatory, 
which is operated by the Association of Universities for Research in Astronomy Inc., under 
cooperative agreement with the National Science Foundation.},
following the normal sequence of preprocessing described on the Hydra webpages.  The raw images were overscan and
bias subtracted, and the detector was flat-fielded with milk flats (daytime sky
exposures) using a diffusing filter in front of the Hydra fibres.  Cosmic rays were subtracted
with a Laplacian edge detection method \citep{vandokkum}.  
Spectral throughput and dispersion corrections were applied based
on the standard continuum lamp and Hydra HeNeAr penray lamp to obtain wavelength calibrated spectra. 

\subsection{Background Light Removal}

Removal of background light from each object spectrum includes
both the uniform dark-sky illumination and (for the GC candidates close to
the galaxy center) the bulge light from NGC 5128 itself.  To trace
both effects, we placed 40-50 fibers in each Hydra field configuration
on blank-sky positions sprinkled across the entire field, then plotted the
mean intensity of each sky fiber versus radius from galaxy center.
The data showed that the bulge light profile was well matched by a $(1/r^2)$ 
radial dependence \citep[see, e.g.][]{hhgII}.

For radii larger than $r \simeq 11'$, the galaxy bulge light contamination was negligible
and we simply subtracted the
average of the sky spectra at these larger radii
from all the object spectra.  
For the inner objects, we then removed the additional bulge light contamination 
through appropriate radial scaling of the 
average bulge sky spectrum. 

\subsection{Radial Velocities}

From the fully preprocessed and background-subtracted spectra, we
derived radial velocities obtained through Fourier cross-correlation, using the 
{\it fxcor} task in the {\it radial velocity} IRAF package.  The spectrum of each
candidate object was correlated against a template spectrum of known velocity and the
correlation peak was fit with a Lorentzian function.  Our template spectrum
was constructed from the sum of five of the previously known bright GCs
in NGC 5128 (C7, C11, C17, C19, and C20), which we observed in both of
our Hydra fields and which have well determined velocities taken from PFF04.
Thus, the template was very well tuned to the the spectral features in
the GCs that we are searching for since it was built from exactly the
same type of object.

\section{Results}

\subsection{Candidate Distribution}
NGC 5128 has a systemic velocity of 541 km s$^{-1}$ \citep{hui} and an
internal stellar velocity dispersion of $\sim 140$ km s$^{-1}$ \citep{wilkinson}.  
These features allow a clean velocity separation between GCs and
foreground stars in the Milky Way or background
galaxies.  As shown (e.g.) by PFF04, the foreground stars 
have velocities almost all less than $\sim 200$ km s$^{-1}$, while any faint background
galaxies will have velocities of many thousands of km s$^{-1}$.  Any objects
that fall
in the range $200 < v_r < 1000$ km s$^{-1}$ (and drawn from the candidate list
pre-selected by image morphology, magnitude and color) will therefore be near-certain GCs.

From both Hydra fields combined, we obtained radial
velocity measurements for 86 candidate GCs and 43 previously
known GCs.  The previously known clusters were used to help establish
the zero point of our velocity scale and check the internal measurement
uncertainties (see below).
Of the candidates, we found 31 with $240 \leq v_r \leq 900$
km s$^{-1}$, which we identify as newly discovered GCs.  Of these 31, 5 are X-ray
sources previously detected by \citet{kraft}.
In Table~\ref{newGC}, we list the data for these GCs.  Successive columns give the
identification number (WHH prefixes denote our new objects), 
coordinates, radial distance from the center of NGC 5128, $V$ and $(C-T_1)$ photometry
from HHG04, radial velocity, and velocity uncertainty from the {\it fxcor} solution.

Of the 55 remaining candidates, 24 are definite foreground stars 
with $-91 \leq v_r \leq 110$ kms$^{-1}$, and 31 are
likely background galaxies; these are listed in Table~\ref{FSBG}. The objects
with null results for the radial velocity are ones whose spectra had a poor correlation 
with the GC template spectrum, suggestive of a background galaxy with either
very large redshift or spectral features very different from the template.  
Confirming 31 candidates as GCs out
of the 86 candidates observed yields an encouraging 36\% hit rate
given the extremely high field contamination level, and demonstrates
the value of rigorous preselection by color and object morphology (see the
discussion of HHG04).  

Table~\ref{knownGC} lists the resulting radial velocity measurements
of the 43 previously known GCs remeasured in this study.  The
identification number prefixes C, G, or pff\_gc are originally listed
by \citet{hghh} and PFF04.

\begin{figure}
\plotone{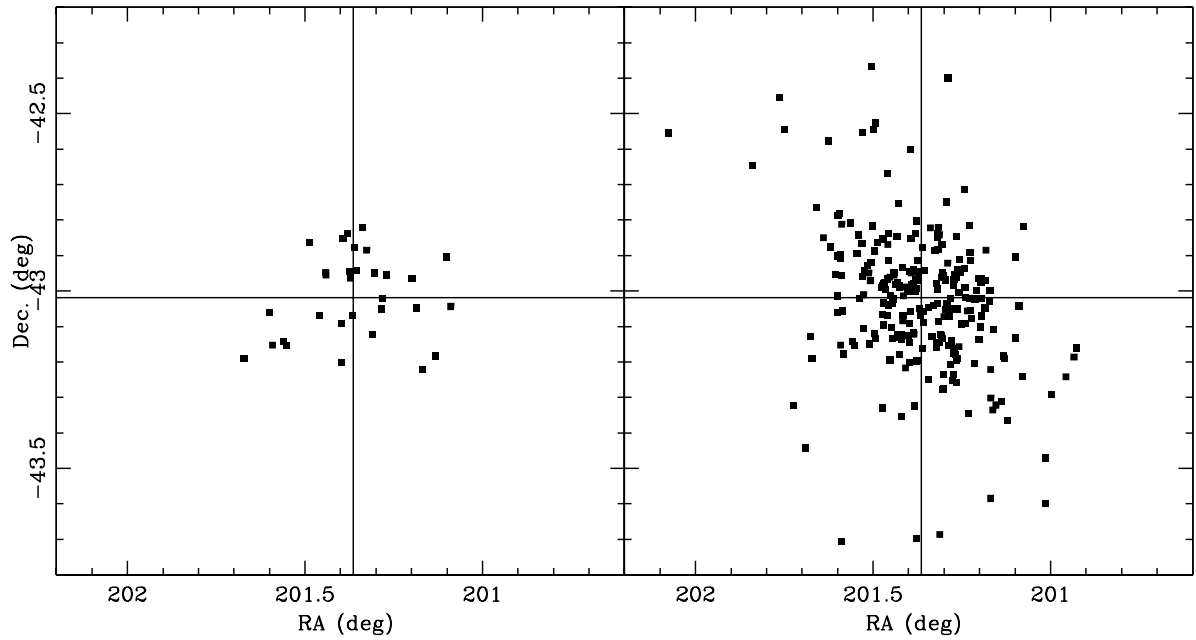}
\caption{The positions in RA and Dec:  ({\it left}) the 31 new globular
  clusters found in this study; ({\it right}) 215 known
  globular clusters from previous literature (Peng, Ford, and Freeman 2004).
  The cross hairs indicate the centre of the galaxy.} 
\label{positions}
\end{figure}

\begin{figure}
\plotone{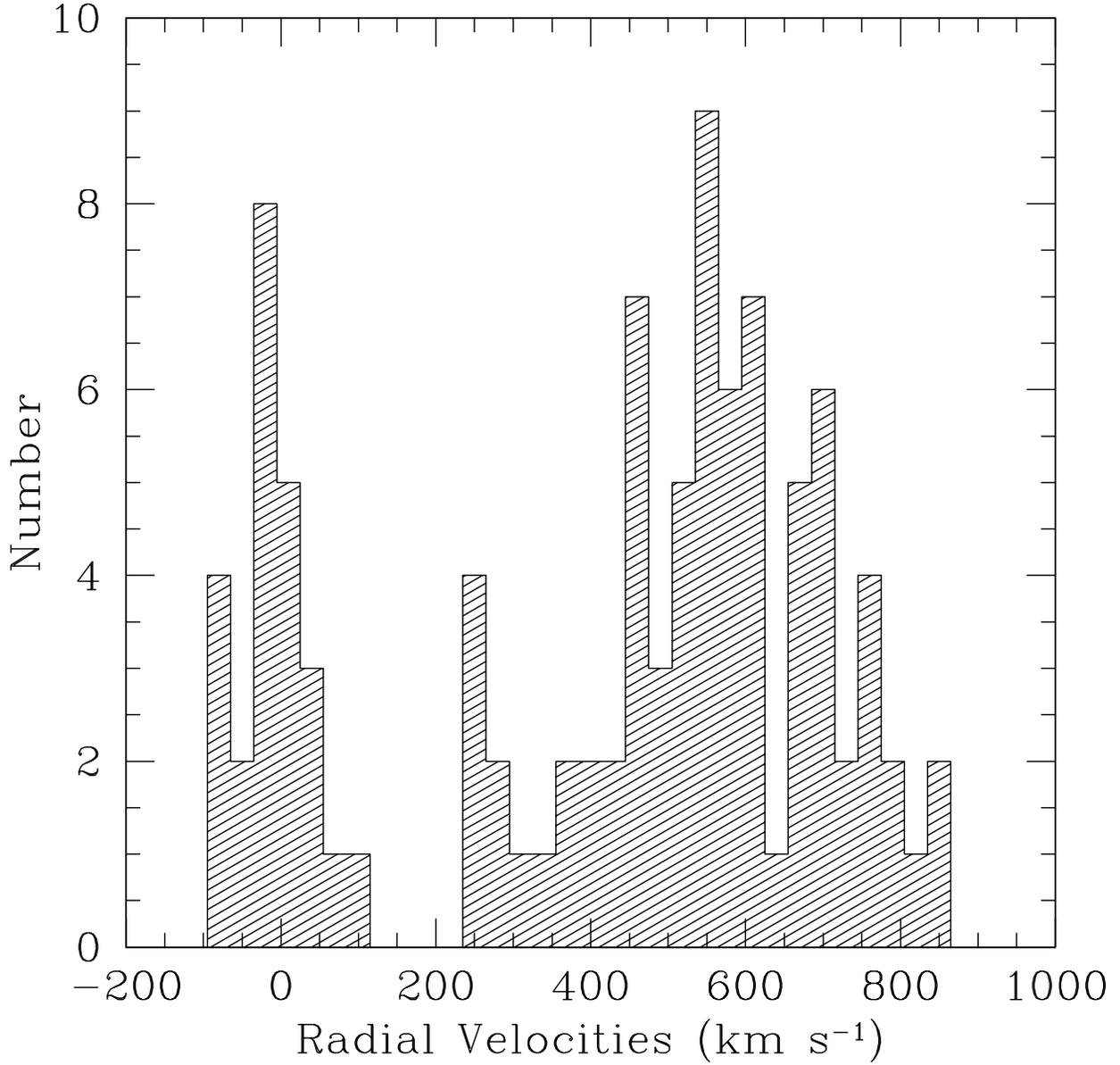}
\caption{The velocity distribution of the 74 measured GCs ($v_r \sim 200$ to $900$
  km s$^{-1}$) and
  foreground stars ($v_r \sim -90$ to $110$ km s$^{-1}$).}
\label{FSGC}
\end{figure}

Figure \ref{positions} shows the locations of the newly
confirmed and previously known GCs.  Whereas most previous surveys have
concentrated on fields along the isophotal major axis of NGC 5128 (along
a line running SW to NE), it is worth noting that
many of our newly identified GCs lie along the
less well studied minor axis parallel to the inner dust lane.
There are likely to be many more to be found still. Figure
\ref{FSGC} shows the velocity distribution of the GCs and foreground 
stars measured from this study; the two types of objects are
clearly distinguished by radial velocity at $v_r \sim 200$ km s$^{-1}$.

The analysis of the GCS radial profile by HHG04 showed that 
the majority of GCs lie within $r = 15'$ from the galaxy 
center, with the main contribution of the clusters residing in an 
even more compact environment within $r = 10'$.  Our new sample of 31
GCs has a median value of $r = 8.25'$.

The mean radial velocity difference between the 
previously known clusters in our
study and those of PFF04 is less than 10 km s$^{-1}$, well
within the standard deviation of the mean (see below).  The zero point 
of our velocity scale therefore needs no adjustment.

\subsection{Comments on Individual Objects}

The raw spectra of clusters WHH-12, WHH-25, WHH-31, pff\_gc-073, and the
non-GC candidates 100, 133, and 136 had a few intensity spikes that
could not be removed in the preprocessing stage.  These regions
were artificially removed before the cross-correlation solutions,
resulting in higher than normal velocity uncertainties 
(see Tables~\ref{newGC} \&~\ref{FSBG}).

For the previously known cluster C10, we measure
$v_r = 855 \pm 33$ km s$^{-1}$.  This differs quite a bit
from the PFF04 measurement of $522 \pm 53$ km s$^{-1}$, 
but it agrees well with \citet{hhh} ($770$ km s$^{-1}$) and
also with the new measurement from \citet{beasley}
(836 km s$^{-1}$).  We therefore believe our measurement to be in
the correct range.

There is one candidate for which the velocity solution
revealed a second correlation peak only slightly smaller than 
the highest peak that is automatically 
found and fit by the {\it fxcor} task.  For WHH-22, the highest correlation peak 
yields $v_r = 492
\pm 126$ km s$^{-1}$, clearly classifying this object as a GC; but a 
second peak yields $v_r \sim 20$ km s$^{-1}$, which would classify it 
as a foreground star.  Visual inspection of the candidate spectrum
shows hydrogen and calcium line strengths consistent with those for
clusters.  We keep this object as a probably GC, but with these
uncertainties in mind.  A strong secondary peak was also revealed for
the known cluster, pff\_gc-060.  The highest peak gives $v_r = 818 \pm
96$ km s$^{-1}$, but a second peak gives 217
km s$^{-1}$, which would put it on the borderline between GCs and
foreground stars.  Its spectrum does not have high enough signal-to-noise to
determine upon visual inspection whether or not it is a cluster.
However, this
object has been measured by PFF04 to have $v_r = 893 \pm
19$ km s$^{-1}$, agreeing with our highest peak correlation value.  

For three previously known GCs, C21, pff\_gc-089, and pff\_gc-064, the
highest correlation peak fit generated by {\sl fxcor}
yielded radial velocities that were inconsistent with 
previous measurements.  A closer look at the 
correlated spectra shows a second strong correlation peak in all cases.
Choosing the secondary peak
gave $v_r = 461 \pm 96$
km s$^{-1}$ for C21 (compared with $465 \pm 7$ km s$^{-1}$ from PFF04)
and $v_r = 553 \pm 103$ km s$^{-1}$ for pff\_gc-089  
(compared with $v_r = 502 \pm 61$ km s$^{-1}$ from PFF04). 
C21 has also been previously measured by
\citet{hhh} to be at $v_r = 490$ km s$^{-1}$.  For these two objects
we therefore adopted the second correlation peaks.  The known cluster,
pff\_gc-064, measured in both field configurations, presents a
slightly different problem:  the initial poorly correlated fit in
field 1 nominally classified it as a possible background galaxy. 
Yet, there is a second strong
correlation peak with $v_r = 548 \pm 97$
km s$^{-1}$, agreeing with the field 2 result for the same
object of $563 \pm 84$ km s$^{-1}$ and also with the PFF04 measurement of $v_r = 557 \pm 31$ km s$^{-1}$.

There are three other candidates measured in both field configurations,
43/44, 109/110, and 101.  Their radial velocities
classified these objects as possible background galaxies, with their
averaged radial velocities as displayed in Tables~\ref{FSBG}.
Objects 43 and 44 in the candidate list of HHG04
turn out to be the same object, and similarly candidates 109 and 110
are the same.  Lastly, we note that 12 of the HHG04 candidate objects measured in this
study turned out to be previously known GCs.  These objects are listed
in Table~\ref{overlaps}.

\subsection{Velocity Measurement Uncertainties}

The radial velocity uncertainties we quote in the tables are 
from the {\it fxcor} solutions.  Generally, these are well correlated
with the object's apparent magnitude and thus the signal-to-noise of the spectra.
Figure~\ref{Resultserrmag3} shows the
nominal radial velocity uncertainties plotted versus V magnitude.
The new GCs (excluding WHH-12, WHH-25, and
WHH-31; see above) with $V < 18.5$ had a mean
$\sigma(v_r) = \pm 32$ km s$^{-1}$ and S/N $ =
13$ per pixel in the continuum region near $\lambda4525 \AA$. For $V \geq 18.5$, 
the mean uncertainty was 
$\sigma(v_r) = \pm 74$ km s$^{-1}$ and S/N $(\lambda 4525 \AA) = 4.5$
per pixel.

\begin{figure}
\plotone{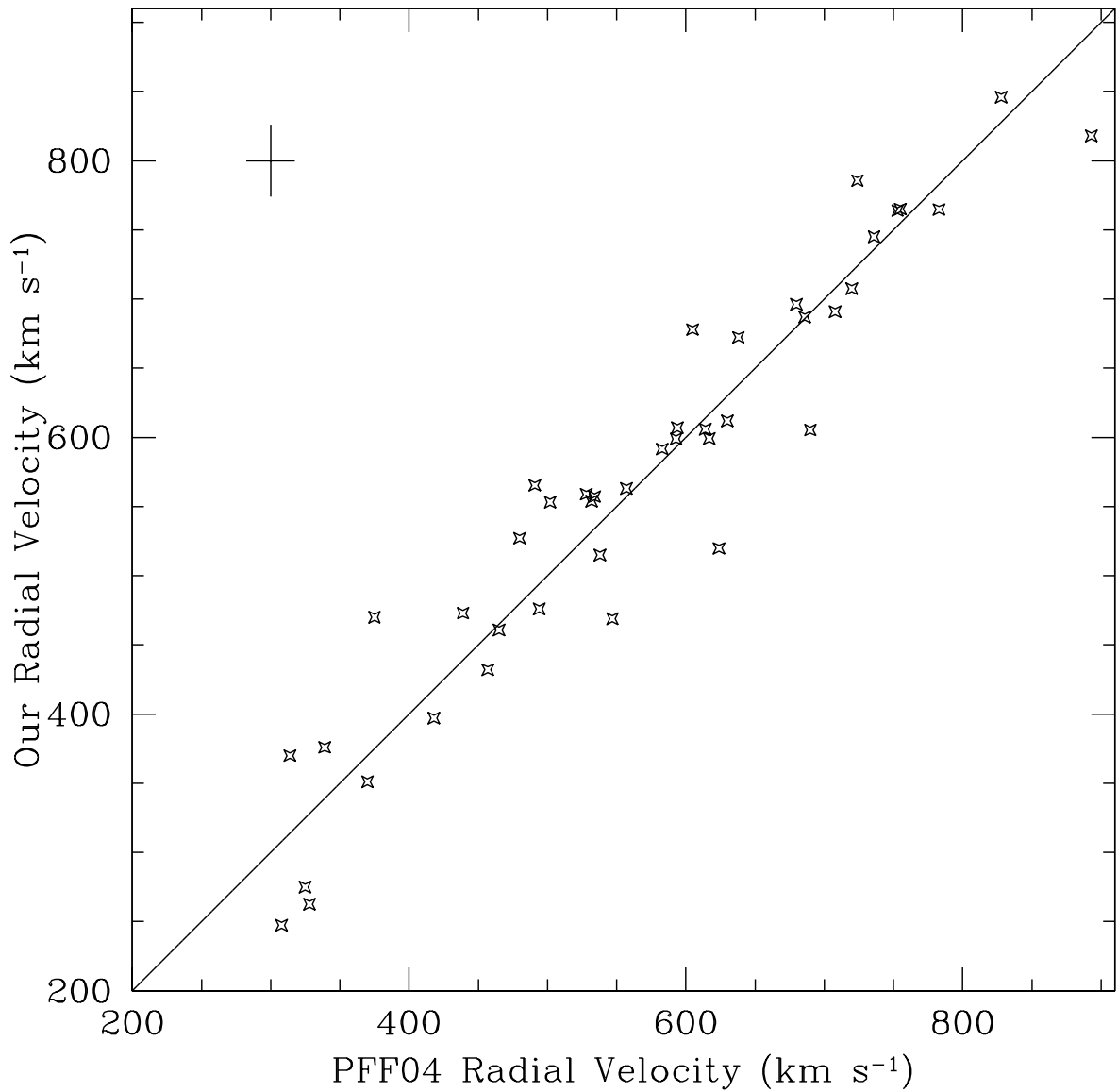}
\caption{Comparison of our radial velocity to the PFF04 radial
  velocity for the 42 known clusters in common (excluding C10; see
  text).  The standard deviation of the differences
 is $\pm$44.5 km s$^{-1}$.  The average
  internal uncertainty of each set of measurements is shown in the
  upper left of the plot.}
\label{comparison}
\end{figure}

\begin{figure}
\plotone{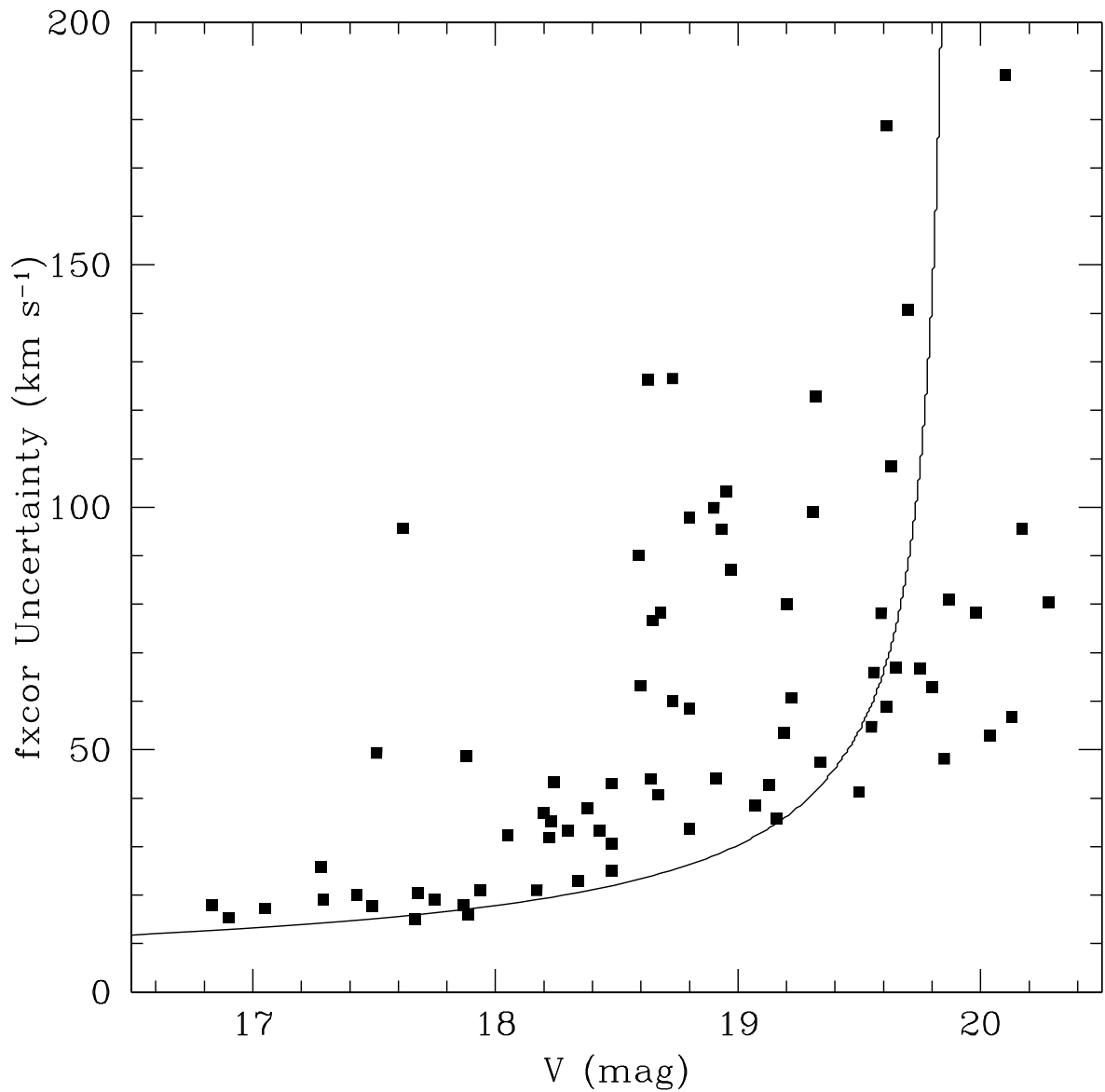}
\caption{The {\sl fxcor} velocity uncertainties obtained for the 74 globular
  clusters listed in Tables 1 \& 3, compared to their V magnitudes. The
  solid line is the ideally expected increase in velocity uncertainty with
  increasing V magnitude, based on the {\it mknoise} simulations.}
\label{Resultserrmag3}
\end{figure}

We tested the {\it fxcor} uncertainties in several ways.
First, the comparison between our velocities and those of
PFF04 for 42 clusters in common (excluding C10 as mentioned above) is shown in 
Figure~\ref{comparison}.  
The standard deviation of the differences between the two studies
was $\pm$44.5 km s$^{-1}$. If the two studies have comparable internal
uncertainties, then this suggests that for GCs brighter than $V \simeq 19$
the typical uncertainty of our measurements is no larger than $\pm 30$ km s$^{-1}$,
in agreement with the {\it mknoise} simulations in
Fig.~\ref{Resultserrmag3} (see below).

Another internal test was to compare the velocities for objects
that we measured in both Hydra field settings.  There were 13 such GCs 
measured twice, 12 of them brighter than $V=19$:  C2, C6, C13, C14,
C18, C19, C20, C26, C37, C38, C44, pff\_gc-056, and pff\_gc-064 ($V =
19.80$ mag).
For these, the mean velocity difference between the two fields was 3.9 km s$^{-1}$,
with a standard deviation of $\pm$22.3 km s$^{-1}$, again consistent with
Fig.~\ref{Resultserrmag3}.

Finally, we compared the {\it fxcor} uncertainties with those generated
from simulated noisy spectra, adding Gaussian noise to the template 
spectrum with the IRAF {\it artdata.mknoise} task.
Several dozen of these artificial spectra 
were cross-correlated against the original template
spectrum and the resulting radial velocity uncertainties plotted against
the S/N for each noisy spectrum.  The results verify that for
S/N $ > 5$ per pixel (corresponding roughly to $V < 18.5$), 
the velocity uncertainty was smaller than $\pm30$ km s$^{-1}$.
Fainter than this, the uncertainty rises exponentially, reaching
$\pm 100$ km s$^{-1}$ at S/N $ \simeq 2$ per pixel (or $V \sim 20$).  In
Fig.~\ref{Resultserrmag3}, the solid line shows the velocity
uncertainty increasing with V magnitude, from the {\it mknoise}
simulations.  This ideal line is the minimum uncertainty
due only to random noise, providing a lower boundary to the observed
uncertainties. 

In summary, we find that the results generated by {\it fxcor} (with
its uncertainities as listed in Tables~\ref{newGC}, \ref{FSBG},
\&~\ref{knownGC}) provide estimates of the internal velocity
uncertainties that are quite close to the estimates from simulations
and from comparisons with previous data.
This may in large part be because we deliberately used a standard
velocity template which accurately matches the program objects.
We conclude that the internal uncertainties of our 
measured velocities are near $\pm 20$ km s$^{-1}$ for the very brightest
clusters, increasing to $\pm 100$ km s$^{-1}$ for the ones at 
$V \sim 20$ or fainter.

\section{The Metallicity Distribution Function}

A new analysis of the radial velocities and
kinematics of the confirmed GCS in NGC 5128, with a much enlarged cluster
sample over that used in PFF04, will be discussed in a
later paper \citep{woodley}.  For the present, we provide here 
only a brief discussion on the metallicity distribution function (MDF).

\begin{figure}
\plotone{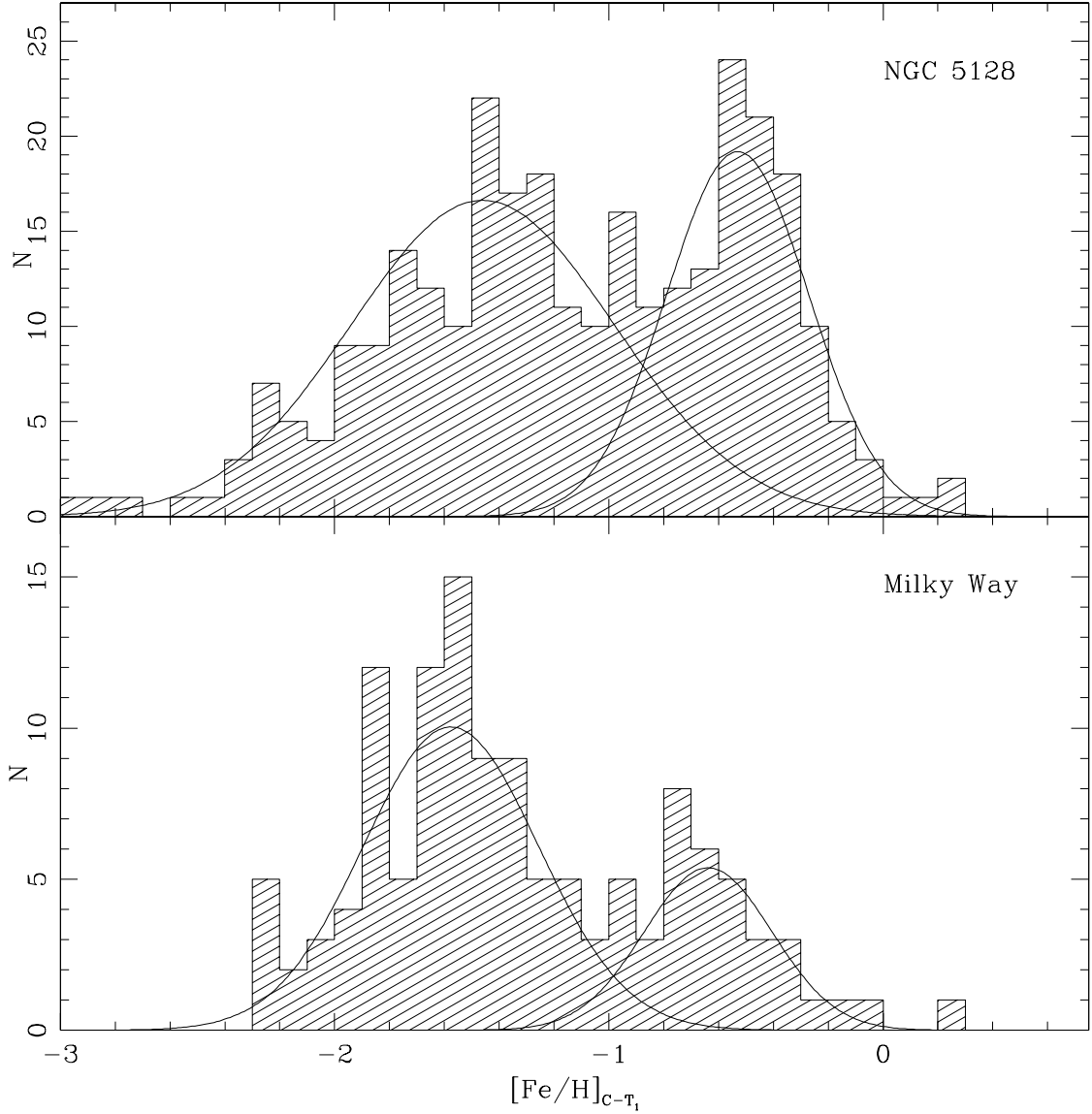}
\caption{{\it Top:} The metallicity distribution function for 299 GCs 
including the 31 new GCs from this survey. The distribution is fit 
with two Gaussians with centroids at [Fe/H] = ($-1.46, -0.53$) and
standard deviations (0.54, 0.30).
{\it Bottom:} The metallicity distribution function
  for 148 Milky Way GCs with Gaussian centroids at [Fe/H] =
  ($-1.58 \pm 0.05$,$-0.64 \pm 0.07$) and standard deviations (0.32,
  0.23)}
\label{mdf}
\end{figure}

The top panel in Figure~\ref{mdf} shows the MDF for a total of 299 GCs in NGC 5128 
that are radial-velocity members {\sl and} that have available $(C-T_1)$ photometric indices:
168 are previously confirmed clusters (PFF04), 31 are the new clusters from this
study, and 100 more are from the upcoming study of \citet{beasley}. 
Although there are a few dozen other highly probable GCs that have recently been found
in NGC 5128
from image morphology studies \citep{rejkuba,har02,gom05}, we restrict the current
discussion to an almost certainly ``pure'' sample generated by the combination of
object color, morphology, and radial velocity.

The dereddened colors $(C-T_1)_0$ from HHG04 were transformed to [Fe/H]  
through the conversion derived by \citet{hh} calibrated through the Milky Way
cluster data.  We adopt a foreground reddening value 
of $\rm{E(B-V)} = 0.11$ for NGC 5128, corresponding to $E(C-T_1) = 0.22$.  Because the transformation 
is slightly nonlinear, the uncertainty in [Fe/H] is a function of metallicity.
For a typical color uncertainty $\sigma(C-T_1) \simeq 0.1$, the corresponding
uncertainty in [Fe/H] is near $\pm 0.2$ dex at the metal-poor end, decreasing
to $\pm 0.07$ dex at the metal-rich end.

The MDF, binned in steps of $\Delta\rm{[Fe/H]} =
0.10$, is shown in Figure \ref{mdf}, along with the same distribution
for the Milky Way clusters \citep[data from][]{har96}.  As has been found
in all previous studies of NGC 5128 starting with \citet{hghh}, the
MDF is extremely broad and populated by roughly equal numbers of
metal-poor and metal-rich clusters.  The color bimodality of NGC 5128 has recently
been confirmed by \citet{peng04} in 4 different photometric indices.  They used spectroscopic line indices of H$\beta$
versus [MgFe]$'$ to suggest the blue GC population was metal-poor
and old ($8-15$ Gyr) and the red GC population was metal-rich and young (mean age $\sim
5^{+3}_{-2}$ Gyr).  A bimodal-Gaussian distribution (shown in Fig.~\ref{mdf}) provides a
reasonable fit to the total MDF, but not as cleanly as in many other
giant galaxies including the Milky Way (plotted for comparison in the lower
panel of the figure).  The fit of the bimodal model to the MDF is not
well constrained, allowing a wide range of particular combinations of
Gaussians.  The two centroids shown, at 
$\rm{[Fe/H]} = -1.46$ and $-0.53$, are similar to the fit obtained by HHG04
(see their Figure~10) from a smaller cluster sample.
The metal-poor part of our entire sample makes up $\sim
54\%$ of the total population and the metal-rich part $\sim 46\%$,
in close agreement with HHG04.   
It does not seem likely that internal differential reddening of the clusters
is an important factor in creating the color spread, since reddening-generated differences in color of
several tenths of a magnitude would be needed to move a cluster from
the metal-poor mode to (e.g.) [Fe/H] $\sim -1$, yet all the objects with $(C-T_1)$ photometry lie
clearly outside the dust lane and into the clear regions of the bulge
and halo \citep[see][for extensive discussion]{hI}.  

As has long been realized, the Milky Way MDF (bottom panel in Figure~\ref{mdf}) provides a clean
match to the bimodal-Gaussian model, 
with peaks at $\rm{[Fe/H]} = -1.58 \pm 0.05$ and $-0.64 \pm 0.07$.  The standard deviations
are 0.32 and 0.23 for the metal-poor and metal-rich subpopulations,
both narrower than their counterparts for NGC 5128.   
Both centroids are slightly but noticeably
more metal-poor than those in NGC 5128.
  
By comparison, the MDF of the M31 globular clusters shows [Fe/H] peaks
that are more similar to NGC 5128.
\citet{perrett} find M31 centroids of $\rm{[Fe/H]} = -1.44 \pm 0.03$
and $ -0.50 \pm 0.04$.  Thus, it is likely that if NGC 5128 is a
product of major mergers, then it would need M31-sized galaxies as
progenitors, rather than systems like the Milky Way which have
lower-metallicity clusters. 

If we assume that NGC 5128 was built through a series of mergers 
extending for several Gigayears, then the GCS might contain 
a wide range of ages.  Thus the MDF shown here, based on photometric
metallicities from $(C-T_1)$ with the assumption that all the clusters
are homogeneously old, would underestimate the true metallicity of 
younger clusters.  This effect would artificially increase the spread in
the MDF and partially blur out a distribution that was intrinsically
bimodal.  The classic age/metallicity degeneracy for broad band colors
prevents us from distinguishing this alternative from an intrinsically
broad, even MDF.  

\begin{figure}
\plotone{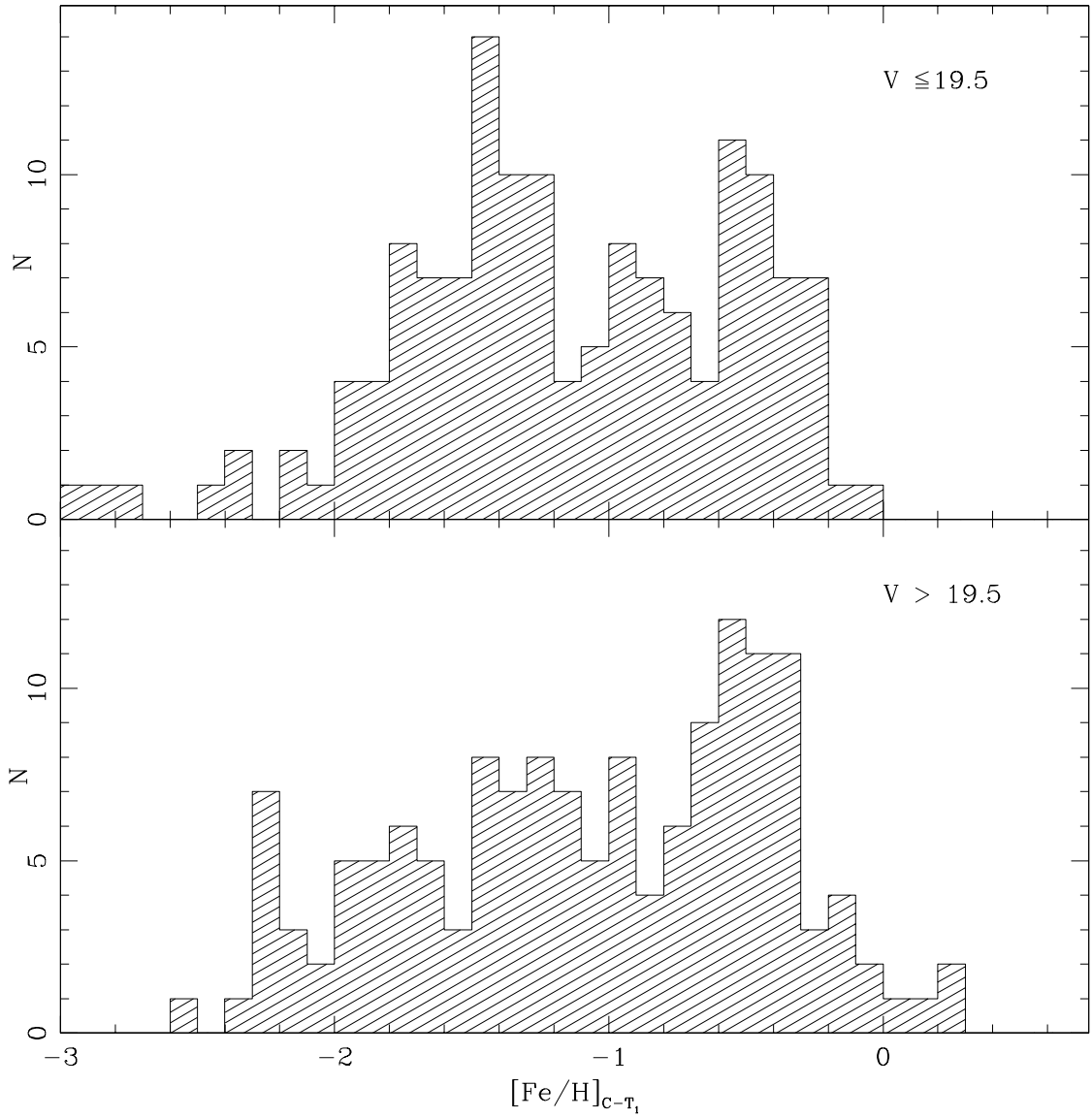}
\caption{The metallicity distributions for the 299 GCs divided in V
  magnitude;  {\it top:} 147 GCs with V $\leq 19.5$ mag and {\it
  bottom:} 152 GCs with V $ > 19.5$ mag.}
\label{fehv}
\end{figure}

Figure \ref{fehv} displays the NGC 5128 sample divided into bright
($V \leq 19.5$) and faint ($V > 19.5$) clusters.
Interestingly, the fainter sample appears to have a higher proportion of
metal-rich clusters than the brighter sample.  A direct least-squares
correlation of [Fe/H] versus V magnitude shows no significant
trend within the larger scatter in [Fe/H].  
It is possible that we are simply seeing an
intrinsically broad and uniform spread of cluster metallicities, in
which the younger and more metal-rich ones tend to be of somewhat
lower mass or luminosity, consistent with the age differences
suggested by PFF04.  A wide range of ages would further increase
the apparent spread of the MDF as seen in Fig.~\ref{fehv}.  

Further spectroscopic studies and more detailed modelling 
will be needed to test these speculations.
In later analyses \citep{woodley,beasley} we will discuss the age
distributions and kinematics of the cluster system further.

\begin{figure}
\plotone{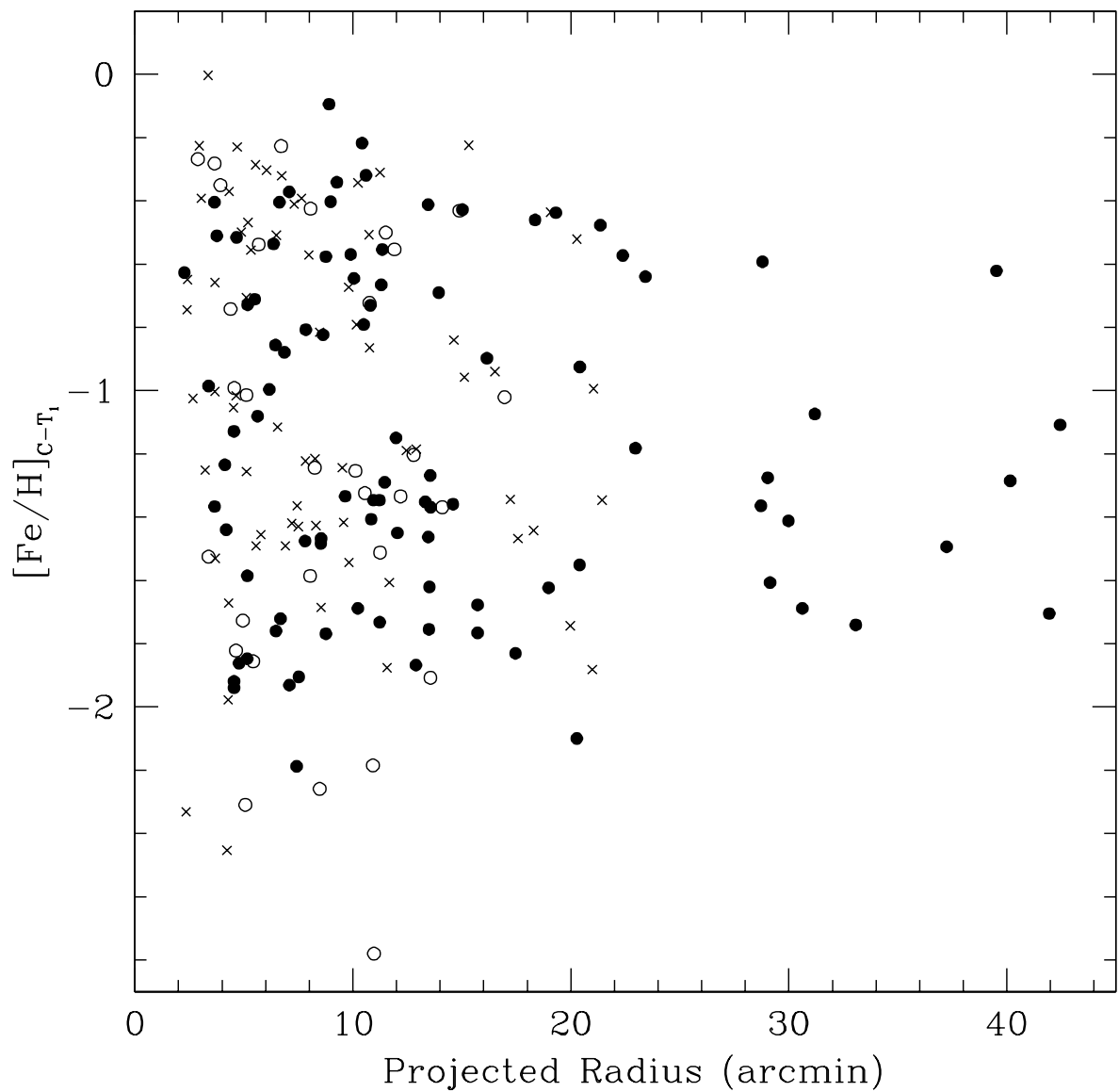}
\caption{Metallicity from the C-T$_1$ colour index vs. projected galactocentric distance for 
  the radial-velocity confirmed GCs. The {\it crosses} denote clusters from
  \citet{hghh,vhh,hhh}, the {\it solid circles} are clusters from
  \citet{pffI}, and the {\it open circles} are new clusters from this study.}
\label{fehrad}
\end{figure}

In Figure \ref{fehrad}, the cluster metallicities are displayed
as a function of projected galactocentric radius.  Both metal-poor ($\rm{[Fe/H]} > -1.0$)
and metal-rich ($\rm{[Fe/H]} < -1.0$) clusters have been found extending out to a radius of $40'$
($\sim 45$ kpc).  However, beyond $r \sim 20'$, there are double the
number of metal-poor compared to metal-rich clusters.  The mean
metallicity beyond this distance is $\rm{[Fe/H]} = -1.21$ with a standard deviation of 0.47. 
As is discussed in more detail in HHG04, this change in ratios of subpopulations is
not likely to be due to selection bias and is therefore probably intrinsic to the system.

\section{Summary}
We present new radial velocities for 86 globular cluster candidates selected from
\cite{hhgII}.  The results show 31 newly identified clusters with $240 \leq v_r \leq 900$
kms$^{-1}$, 24 definite foreground stars with $-91 \leq v_r \leq 110$
km s$^{-1}$, and 31 probable background galaxies.  We also obtained velocities
for 43 previously known GCs in NGC 5128.
The internal uncertainty in our $v_r$ measurements ranges from $\pm 20$ km s$^{-1}$
at the bright end of the sample ($V < 18)$ to $\sim \pm 100$ km s$^{-1}$ at $V \sim 20$.

Combining our data with previously published material, we derive a
metallicity distribution for the GCs based on 
299 clusters.  A bimodal distribution
with centroids at $\rm{[Fe/H]} = -1.46$ and $-0.53$ provides an approximate fit
to the MDF, though not highly accurate, and the data hint that the
relative proportions of metal-poor and metal-rich clusters may 
change with luminosity.  Two possible interpretations are that (a) we may be seeing the 
combined effects of a several-Gy internal age spread in
the cluster system, combined with a trend for the younger 
clusters to be progressively less massive, or (b) the MDF is genuinely
broad and evenly distributed, unlike the more cleanly bimodal MDFs
seen in other giant galaxies.

\acknowledgments
We greatly appreciate the data reduction advice and support provided
by Knut Olsen at CTIO, and also advice regarding
uncertainty tests from David James at Vanderbilt University.
We thank Mike Beasley for transmitting cluster velocity data in
advance of publication.
This work was supported by the Natural Sciences and Engineering
Research Council of Canada through research grants to GLHH and WEH.




\clearpage

\begin{deluxetable}{rccccccccl}
\tablecaption{Radial Velocity Measurements for Globular Clusters \label{newGC}}
\tablewidth{0pt}
\tablehead{
\colhead{Cluster} & \colhead{RA} & \colhead{Dec} & \colhead{R$_{gc}$} & \colhead{V} & \colhead{(C-T$_1$)}  &
\colhead{$v_r$}  & \colhead{$\sigma(v_r)$} \\
\colhead{ } &\colhead{(J2000)} & \colhead{(J2000)}
&\colhead{(arcmin)}& \colhead{(mag)}&\colhead{ }  & \colhead{(km s$^{-1}$)} & \colhead{(km s$^{-1}$)} \\
}

\startdata
WHH-1 & 13 24 21.40 & -43 02 36.8 & 12.19 & 18.23 & 1.484 & 600 &  35  \\
WHH-2 & 13 24 23.98 & -42 54 10.7 & 13.56 & 19.87 & 1.270 & 582 &  81  \\
WHH-3 & 13 24 32.17 & -43 10 56.9 & 14.10 & 19.59 & 1.470 & 636 &  78  \\
WHH-4 & 13 24 40.60 & -43 13 18.1 & 14.89 & 19.13 & 1.927 & 685 &  43  \\
WHH-5 & 13 24 44.58 & -43 02 47.3 &  8.04 & 19.50 & 1.386 & 671 &  41  \\
WHH-6 & 13 24 47.37 & -42 57 51.2 &  8.06 & 19.75 & 1.931 & 712 &  67  \\
WHH-7\tablenotemark{a} & 13 25 05.02 & -42 57 15.0 &  5.68 & 17.43 & 1.863 & 722 &  20  \\
WHH-8\tablenotemark{a} & 13 25 07.62 & -43 01 15.2 &  3.66 & 18.05 & 2.026 & 690 &  32 \\
WHH-9 & 13 25 08.51 & -43 02 57.4 &  3.93 & 18.90 & 1.979 & 315 & 100 \\
WHH-10 & 13 25 12.84 & -42 56 59.8 &  4.95 & 19.70 & 1.334 & 664 & 141 \\
WHH-11\tablenotemark{a} & 13 25 14.24 & -43 07 23.5 &  6.71 & 19.55 & 2.066 & 562 &  55 \\
WHH-12 & 13 25 18.27 & -42 53 04.8 &  8.25 & 19.31 & 1.521 & 558 &  99\tablenotemark{b} \\
WHH-13 & 13 25 21.29 & -42 49 17.7 & 11.91 & 19.34 & 1.854 & 467 &  47 \\
WHH-14 & 13 25 25.49 & -42 56 31.2 &  4.64 & 20.28 & 1.300 & 436 &  80 \\
WHH-15 & 13 25 26.78 & -42 52 39.9 &  8.48 & 20.13 & 1.153 & 506 &  57 \\
WHH-16\tablenotemark{a} & 13 25 27.97 & -43 04 02.2 &  2.89 & 19.20 & 2.036 & 749 &  80 \\
WHH-17 & 13 25 29.25 & -42 57 47.1 &  3.38 & 18.73 & 1.409 & 588 & 127 \\
WHH-18 & 13 25 30.07 & -42 56 46.9 &  4.39 & 18.48 & 1.752 & 771 &  31 \\
WHH-19 & 13 25 31.03 & -42 50 14.9 & 10.92 & 17.51 & 1.177 & 470 &  49 \\
WHH-20 & 13 25 34.36 & -42 51 05.9 & 10.12 & 19.19 & 1.517 & 242 &  54 \\
WHH-21 & 13 25 35.22 & -43 12 01.5 & 10.97 & 19.22 & 0.994 & 243 &  61 \\
WHH-22\tablenotemark{a} & 13 25 35.31 & -43 05 29.0 &  4.56 & 18.63 & 1.631 & 492\tablenotemark{c} & 126 \\
WHH-23 & 13 25 45.90 & -42 57 20.2 &  5.07 & 18.60 & 1.137 & 286 &  63 \\
WHH-24 & 13 25 46.00 & -42 56 53.0 &  5.43 & 19.85 & 1.288 & 566 &  48 \\
WHH-25 & 13 25 50.34 & -43 04 08.2 &  5.12 & 20.17 & 1.621 & 551 &  95\tablenotemark{b} \\
WHH-26 & 13 25 56.59 & -42 51 46.6 & 10.76 & 19.16 & 1.762 & 412 &  36 \\
WHH-27 & 13 26 12.82 & -43 09 09.2 & 11.51 & 18.73 & 1.885 & 545 &  60 \\
WHH-28 & 13 26 14.18 & -43 08 30.4 & 11.25 & 19.32 & 1.414 & 506 & 123 \\
WHH-29 & 13 26 22.08 & -43 09 10.7 & 12.79 & 19.98 & 1.538 & 505 &  78 \\
WHH-30 & 13 26 23.60 & -43 03 43.9 & 10.55 & 19.56 & 1.488 & 470 &  66 \\
WHH-31 & 13 26 41.43 & -43 11 25.0 & 16.96 & 19.65 & 1.618 & 573 &  67\tablenotemark{b} \\
\enddata

\tablenotetext{a}{Detected as X-ray source from \citet{kraft}}
\tablenotetext{b}{Regions of spectrum affected by cosmic ray spikes were masked out during correlation.}
\tablenotetext{c}{Strong secondary correlation peaks.}

\end{deluxetable}

\clearpage

\begin{deluxetable}{rccccccc}
\tablecaption{Foreground Stars and Background Galaxies \label{FSBG}}
\tablewidth{0pt}
\tablehead{
\colhead{ID} & \colhead{R.A.} & \colhead{Dec.}  & 
\colhead{$v_r$}  & \colhead{$\sigma(v_r)$}  & \colhead{Comments}\\
\colhead{ } &\colhead{(J2000)} & \colhead{(J2000)} & \colhead{(km s$^{-1}$)} &
\colhead{(km s$^{-1}$)} &\colhead{ }  \\
}

\startdata
36	 & 13 24 02.55 & -42 49 29.4  &   64	&       42 & 	star\\
3	 & 13 24 28.96 & -43 10 58.3  &  -24	& 	23 & 	star\\
8	 & 13 24 52.11 & -43 08 24.3  &   -1	& 	26 & 	star\\
19	 & 13 24 56.46 & -43 06 34.7  &  -46	& 	24 & 	star\\
11	 & 13 25 04.20 & -43 02 29.3  &  -30	& 	26 & 	star\\
14	 & 13 25 08.59 & -43 07 43.4  &  -17	& 	47 & 	star\\
13	 & 13 25 10.06 & -43 07 40.7  &  -30	& 	21 & 	star\\
26	 & 13 25 17.79 & -43 13 05.6  &  -70	& 	42 & 	star\\
25	 & 13 25 24.37 & -43 12 23.1  &  -13	& 	23 & 	star\\
62	 & 13 25 25.10 & -43 12 27.0  &  -82	& 	79 & 	star\\
27	 & 13 25 42.62 & -43 10 21.4  &  -20	& 	32 & 	star\\
77	 & 13 25 47.98 & -43 00 57.6  &  106	& 	106& 	star\\
4	 & 13 25 55.77 & -43 03 00.9  &   18	& 	21 & 	star\\
1	 & 13 25 56.40 & -42 59 37.3  &   40	& 	23 & 	star\\
2	 & 13 25 57.38 & -42 59 48.5  &    1	& 	22 & 	star\\
6	 & 13 25 58.39 & -42 55 27.6  &  -22	& 	21 & 	star\\
23	 & 13 25 58.44 & -43 03 06.4  &   -3	& 	25 & 	star\\
20	 & 13 26 10.41 & -43 00 53.6  &  -11	& 	26 &    star\\
73	 & 13 26 15.41 & -42 57 46.8  &    3	& 	118& 	star\\
56	 & 13 26 17.81 & -43 00 29.7  &  -60	& 	37 & 	star\\
17	 & 13 26 20.74 & -43 09 13.2  &   51	& 	33 & 	star\\
89	 & 13 26 21.11 & -43 03 08.2  &  -85	& 	99 & 	star\\
61	 & 13 26 32.79 & -43 02 59.4  &   31	& 	50 & 	star\\
38	 & 13 26 33.08 & -43 03 14.0  &  -91	& 	40 & 	star\\
283	 & 13 24 04.50 & -42 48 16.2  & 47636	&        81&    galaxy\\
285	 & 13 24 05.49 & -42 47 24.1  & 4729	& 	124& 	galaxy\\
290	 & 13 24 13.26 & -42 57 42.1  &   -	& 	-  & 	PC\tablenotemark{c}\\
94	 & 13 24 28.15 & -42 53 04.6  & 35382	&       178&    galaxy\\
67	 & 13 24 28.46 & -43 14 56.3  &   -	& 	-  & 	PC\\
122	 & 13 24 29.70 & -43 02 06.4  &   -	& 	-  & 	PC\\
43/44    & 13 24 30.86 & -42 52 40.8  & 35527\tablenotemark{a}& 70&galaxy\\
117	 & 13 24 34.38 & -42 57 32.4  &   -	& 	-  & 	PC\\
127	 & 13 24 34.61 & -43 12 50.5  & 7717	& 	194&    galaxy\\
133	 & 13 24 35.30 & -43 06 15.4  &   -	& 	-\tablenotemark{b}& PC\\
119	 & 13 24 48.71 & -42 52 35.5  &   -	& 	-  & 	PC\\
298	 & 13 25 01.09 & -42 48 21.9  &   -	& 	-  & 	PC\\
136	 & 13 25 01.16 & -43 11 59.0  &   -	& 	-\tablenotemark{b}& PC\\
299	 & 13 25 01.29 & -42 54 26.5  & 2544	& 	92 &    galaxy\\
132	 & 13 25 23.45 & -42 53 26.2  & 2017	& 	78 &    galaxy\\
108	 & 13 25 25.73 & -43 05 16.6  &   -	& 	-  & 	PC\\
103	 & 13 25 28.72 & -42 50 12.3  &   -	& 	-  & 	PC\\
314	 & 13 25 37.53 & -42 55 33.5  &   -	& 	-  & 	PC\\
116	 & 13 25 39.63 & -43 04 01.4  & 32124	& 	253&    galaxy\\
123	 & 13 25 41.37 & -43 12 12.6  & 31018	& 	230&    galaxy\\
317	 & 13 25 45.63 & -43 01 15.5  & 5850	& 	91 &    galaxy\\
72	 & 13 25 49.23 & -43 00 02.2  & 7349	& 	87 &    galaxy\\
90	 & 13 25 49.26 & -43 02 20.4  & 9942	& 	113&    galaxy\\
135	 & 13 25 49.95 & -43 02 26.3  & 38298	& 	114&    galaxy\\
107	 & 13 25 56.63 & -42 58 45.5  &   -	& 	-&      PC\\
101	 & 13 26 02.24 & -43 08 55.7  &   -\tablenotemark{a}& -& PC\\
109/110  & 13 26 02.23 & -43 17 34.8  &   -\tablenotemark{a}& -  & PC\\
134	 & 13 26 17.27 & -43 06 39.3  & 24742	& 	79 &    galaxy\\
113	 & 13 26 21.29 & -42 57 19.1  &   -	& 	-  & 	PC\\
125	 & 13 26 23.64 & -43 00 45.6  &   -	& 	-  & 	PC\\
100	 & 13 26 50.44 & -42 50 36.4  &   -	& 	-\tablenotemark{b} & PC\\
\enddata

\tablenotetext{a}{Measured in both fields; the radial velocity is the average.}
\tablenotetext{b}{Regions of spectrum ignored during correlation to remove spikes, leading to high uncertainties.}
\tablenotetext{c}{PC denotes a poor correlation of the object to the GC template spectrum.}

\end{deluxetable}

\clearpage

\begin{deluxetable}{rccccccccl}
\tablecaption{Radial Velocity Measurements for Previously Known Globular Clusters \label{knownGC}}
\tablewidth{0pt}
\tablehead{
\colhead{Cluster} & \colhead{RA} & \colhead{Dec.} & \colhead{R$_{gc}$} & \colhead{$V$} & \colhead{(C-T$_1$)}  &
\colhead{$v_r$}  & \colhead{$\sigma(v_r)$} \\
\colhead{ } &\colhead{(J2000)} & \colhead{(J2000)}
&\colhead{(arcmin)}& \colhead{(mag)}&\colhead{ }  & \colhead{(km s$^{-1}$)} & \colhead{(km s$^{-1}$)} \\
}

\startdata
C2	   & 13 24 51.47  & -43 12 11.2 & 12.86 & 18.38 & 1.546 &  606\tablenotemark{a} &  38   \\ 
C3         & 13 24 58.21  & -42 56 10.0 &  7.33 & 17.68 & 1.940 &  559 &  20  \\
C4	   & 13 25 01.81  & -43 09 25.5 &  9.53 & 17.88 & 1.451 &  687 &  49 \\
C6	   & 13 25 22.19  & -43 02 45.6 &  1.89 & 16.83 &   -   &  846\tablenotemark{a} &  18  \\
C7	   & 13 26 05.40  & -42 56 32.4 &  8.30 & 16.90 & 1.533 &  599 &  15 \\ 
C10	   & 13 24 48.04  & -43 08 14.3 & 10.13 & 18.30 & 1.727 &  856 &  33   \\
C11	   & 13 24 54.70  & -43 01 21.6 &  6.02 & 17.67 & 2.011 &  765 &  15 \\  
C13	   & 13 25 06.22  & -43 15 11.6 & 14.58 & 18.48 & 1.703 &  599\tablenotemark{a} &  25 	\\
C14	   & 13 25 10.49  & -42 44 52.8 & 16.56 & 17.75 & 1.655 &  691\tablenotemark{a} &  19 	\\
C15	   & 13 25 30.39  & -43 11 49.6 & 10.69 & 18.43 & 1.881 &  672 &  33 	\\
C17	   & 13 25 39.72  & -42 55 59.1 &  5.62 & 17.49 & 1.422 &  785 &  18 \\
C18	   & 13 25 39.86  & -43 05 01.8 &  4.48 & 17.29 & 1.603 &  476\tablenotemark{a} &  19   \\
C19	   & 13 25 43.38  & -43 07 22.9 &  6.87 & 17.87 & 0.324 &  607\tablenotemark{a} &  18 \\
C20	   & 13 25 49.68  & -42 54 49.3 &  7.50 & 17.89 & 1.472 &  745\tablenotemark{a} &  16 \\
C21	   & 13 25 52.73  & -43 05 46.5 &  6.52 & 17.62 & 1.576 &  461 &  96 \\
C26	   & 13 26 15.25  & -42 48 29.4 & 15.36 & 17.94 & 2.068 &  370\tablenotemark{a} &  21 	\\
C30	   & 13 24 54.35  & -42 53 24.7 &  9.84 & 17.05 & 1.788 &  786 &  17   \\
C32	   & 13 25 03.37  & -42 50 46.1 & 11.29 & 18.22 & 2.006 &  708 &  32 	\\
C33	   & 13 25 16.26  & -42 50 53.3 & 10.47 & 18.67 &   -	 &  520 &  41 	\\
C36	   & 13 26 07.73  & -42 52 00.2 & 11.72 & 18.24 & 1.378 &  696 &  43 \\
C37	   & 13 26 10.57  & -42 53 42.7 & 10.81 & 18.17 & 1.691 &  612\tablenotemark{a} &  21 \\
C38	   & 13 26 23.77  & -42 54 01.1 & 12.50 & 18.34 & 1.544 &  397\tablenotemark{a} &  23   \\ 
C39	   & 13 26 42.00  & -43 07 44.9 & 15.11 & 17.28 & 1.647 &  247 &  26 	\\
C44	   & 13 25 31.73  & -43 19 22.8 & 18.25 & 18.48 & 1.441 &  515\tablenotemark{a} &  43 \\
C47	   & 13 25 49.94  & -42 52 09.3 &  9.88 & 18.80 & 1.402 &  592 &  34 \\
C48	   & 13 25 49.83  & -42 50 15.1 & 11.63 & 18.65 & 1.281 &  557 &  77\\
C50	   & 13 26 19.64  & -43 03 18.6 &  9.75 & 18.59 & 1.633 &  605 &  90  \\
G271	   & 13 25 13.94  & -42 57 42.5 &  4.25 & 18.80 & 1.092 &  263 &  98 	\\
G342	   & 13 25 05.75  & -42 59 00.3 &  4.53 & 18.20 &   -   &  565 &  37 	\\
G369	   & 13 24 57.49  & -42 59 23.3 &  5.78 & 18.80 & 1.436 &  527 &  58 \\
pff\_gc-018 & 13 24 47.08  & -43 06 01.7 &  8.87 & 18.91 & 1.603 &  554 &  44 	\\
pff\_gc-041 & 13 25 11.14  & -43 03 09.6 &  3.62 & 19.61 & 1.471 &  432 & 179 \\
pff\_gc-042 & 13 25 12.19  & -43 16 33.7 & 15.67 & 18.97 & 1.352 &  678 &  87   \\
pff\_gc-056 & 13 25 32.80  & -42 56 24.3 &  4.84 & 18.64 & 1.286 &  275\tablenotemark{a} &  44 	\\
pff\_gc-058 & 13 25 35.12  & -42 56 45.1 &  4.60 & 18.68 & 1.259 &  376 &  78 	\\
pff\_gc-060 & 13 25 42.42  & -42 59 02.5 &  3.43 & 18.93 & 1.634 & 818\tablenotemark{c} &  96 \\
pff\_gc-064 & 13 25 43.89 & -42 50 42.5 & 10.85 & 19.80 & 1.758 &556\tablenotemark{a,c}&  63\\
pff\_gc-073 & 13 25 52.76  & -42 58 41.6 &  5.21 & 20.10 & 1.759 & 470 & 189\tablenotemark{b} \\
pff\_gc-075 & 13 25 53.47  & -43 03 56.6 &  5.49 & 19.61 & 1.768 & 764 &  59 \\
pff\_gc-076 & 13 25 53.74  & -43 19 48.5 & 19.26 & 19.07 & 1.923 &  351 &  39 	\\
pff\_gc-083 & 13 26 01.81  & -42 58 15.0 &  6.89 & 20.04 & 1.683 & 473 &  53 \\
pff\_gc-085 & 13 26 06.39 & -43 00 38.1 &  7.11 & 19.63 & 1.262 & 469 & 108 \\
pff\_gc-089 & 13 26 20.19  & -43 10 35.7 & 13.48 & 18.95 & 1.324 &  553 & 103 	\\	     
\enddata
\tablenotetext{a}{Measured in both fields; the radial velocity listed
  is the average.}
\tablenotetext{b}{Regions of spectrum affected by cosmic ray spikes were masked out during correlation.}
\tablenotetext{c}{Strong secondary correlation peaks.}

\end{deluxetable}

\begin{deluxetable}{rr}
\tablecaption{Previously Known Clusters in HHG04 Candidate List\label{overlaps}}
\tablewidth{0pt}
\tablehead{
\colhead{Candidate ID} & \colhead{Cluster}\\
}

\startdata
9    & C18         \\
12   & C3          \\ 
16   & C14         \\
30   & C2          \\ 
42   & pff\_gc-056 \\
48   & pff\_gc-060 \\
82   & pff\_gc-041 \\
83   & pff\_gc-075 \\
85   & pff\_gc-085 \\
95   & pff\_gc-064 \\
111  & pff\_gc-083 \\
118  & pff\_gc-073 \\
\enddata
\end{deluxetable}



\clearpage

\clearpage

\clearpage

\clearpage

\clearpage

\clearpage

\clearpage

\end{document}